\documentstyle[aps,twocolumn,prb,epsf]{revtex}

\begin{document}

\draft

\twocolumn[\hsize\textwidth\columnwidth\hsize\csname @twocolumnfalse\endcsname

\title{A new perturbation treatment applied to the
       transport through a quantum dot}

\author{Luis Craco and Kicheon Kang\cite{kang}}

\address{   Max-Planck-Institut f\"ur Physik Komplexer Systeme,
            N\"othnitzer Strasse 38, D-01187 Dresden, Germany }

\date{\today}

\maketitle

\widetext

\begin{abstract}
\noindent

Resonant tunnelling through an Anderson impurity is investigated by
employing a new perturbation scheme at nonequilibrium. This new approach 
gives the correct weak and strong coupling limit in $U$ by introducing
adjustable parameters in the self-energy and imposing
self-consistency of the occupation number of the impurity. We 
have found that the zero-temperature linear response conductance 
agrees well with that obtained from the exact sum rule. At finite 
temperature the conductance shows a nonzero minimum at the Kondo 
valley, as shown in recent experiments. The effects of an applied bias 
voltage on the single-particle density of states and on the differential 
conductances are discussed for Kondo and non-Kondo systems. 
\end{abstract}

]

\narrowtext

Electronic transport through an artificially fabricated quantum dot (QD) is 
of considerable current interest (for a review see e.g. Ref.\onlinecite{kouwen97} 
and references therein). In addition to the Coulomb blockade and ordinary 
resonant tunnelling through a discrete electronic level, a novel Kondo-assisted
tunnelling has been predicted.~\cite{glazman88,ng88,hershfeld,meir,yeyati93}
This phenomenon has been verified 
in single-electron transistors (SET)~\cite{gordon,cronen98}
as well as in nanometer-sized metallic contacts.~\cite{ralph94}
In the SET structures several parameters can be controlled by external gates.
This opens the possibility of studying
a variety of electron correlation effects 
which are not available in the traditional Kondo problem of dilute magnetic
impurities in a metal host. For example, the level position of the quantum dot 
and the coupling to reservoirs can be tuned in the single electron transistor 
structure~\cite{gordon,cronen98}, so that a {\em tunable Kondo effect} could 
be observed. Further, by applying an external voltage between two reservoirs, 
it is possible to study a new kind of Kondo effect out of equilibrium.

Transport through an Anderson impurity out of equilibrium has been studied 
theoretically by different methods including the perturbation treatment in 
Coulomb repulsion $U$~\cite{hershfeld,yeyati93} and the non-crossing 
approximation (NCA)~\cite{meir}. The second order perturbation treatment 
gives reliable results for the symmetric case (\mbox{$2\varepsilon_0
+U=0$}, $\varepsilon_0$ being the level position of the impurity). However, 
it is known that away from the symmetric case second order 
treatment does not reproduce the correct low and high energy 
limits.~\cite{yeyati93} For the limit of infinite $U$ with weak 
hybridization, NCA provides a quantitatively accurate results for the Kondo 
physics except for very small temperatures. Meanwhile, NCA has some 
drawbacks in the low-energy region: it fails to satisfy Fermi-liquid 
relations~\cite{kuramoto84} and, as a consequence, overestimates the 
conductance due to the Kondo resonance~\cite{meir} at low temperature. 
Although the qualitative 
physics of electron transport through the Anderson impurity is now well 
understood, a theory which can describe properly the wide range of 
parameters, including the physics of Kondo resonant transport, charge 
fluctuation and the empty site limit, is still missing.

In this paper, we investigate the transport through an Anderson impurity
by means of a new perturbation treatment which gives the correct weak 
and strong coupling limits in a self-consistent manner. Therefore, 
one can deal with an interpolation scheme between two limits.
This treatment has been suggested by Kajueter and Kotliar~\cite{kajueter96} 
for an application to the Hubbard model in the high dimension limit. We 
extend this idea to study the transport through an Anderson impurity
in a non-equilibrium regime. We have found that the zero-temperature 
linear-response conductance obtained from the density of states agrees 
well with those obtained from Langreth's exact relation.~\cite{langreth66} 
Such good agreement has not been obtained before by the previous 
theories. Starting from this good agreement, we discuss the effects of 
finite temperature and of finite voltage on transport.

%The effect of finite temperature and voltage are discussed.

The resonant tunnelling through a single quantum state can
be described by the Anderson impurity model 
\begin{eqnarray}
 {\cal H} &=& \sum_{k,\sigma,\alpha}\varepsilon_k^\alpha 
            c_{k\sigma\alpha}^\dagger c_{k\sigma\alpha}
   + \sum_\sigma \varepsilon_0 d_\sigma^\dagger d_\sigma
   + U \hat{n}_\uparrow \hat{n}_\downarrow \nonumber \\
  &+& \sum_{k,\sigma,\alpha} \left( t_{k}^{\alpha}c_{k\sigma\alpha}^\dagger
       d_\sigma + \mbox{\rm H.C.} \right) ,
\end{eqnarray}
where $\varepsilon_k^\alpha$ represents the single particle energy
in the reservoir $\alpha(=L,R)$ with their chemical potential difference
being the applied voltage, that is $\mu_L-\mu_R=eV$.
The parameters $\varepsilon_0$, $U$, and $t_k^\alpha$ denote the
single particle energy in the QD, the Coulomb repulsion, and the
coupling between QD and reservoir states, respectively.

In the wide-band limit of the reservoirs, the current formula can 
be written in the form~\cite{meir92,kang98}
\begin{equation}
 I = \frac{2e}{\hbar} \sum_{\sigma}\int d\omega\, 
     \tilde{\Gamma}(\omega)
     \left\{ f_L(\omega)-f_R(\omega) \right\} \rho_{\sigma}(\omega) ,
               \label{eq:curr}
\end{equation}
where $\tilde{\Gamma}(\omega) = \Gamma_L(\omega) \Gamma_R(\omega) /
 \Gamma(\omega)  $
with $\Gamma_\alpha(\omega) = \pi\sum_k| t_k^\alpha |^2\,\delta(\omega
-\varepsilon_{k\alpha})$ being the coupling strength between the 
QD level and the lead $\alpha$, and $\Gamma(\omega) = \Gamma_L(\omega)
+\Gamma_R(\omega)$.
$f_\alpha(\omega)=1/( e^{\beta(\omega-\mu_\alpha)}+1 )$ 
and $\rho_{\sigma}(\omega)=-\frac{1}{\pi}
\mbox{Im}\,G_{\sigma}(\omega)$
are the Fermi function of lead $\alpha$ 
and the spectral density of states  (DOS) of the electron in the
QD, respectively.
  
In calculating the Green's function 
\begin{equation}
\label{eq:Gf}
G_\sigma(\omega) = \frac{1}{\omega - \varepsilon_0 
- \sum_{\alpha} \Delta_\alpha (\omega)- \Sigma(\omega)}\;,
\end{equation}
$\Delta_\alpha(\omega) = \sum_{k\in\alpha} 
|t_{k}^{\alpha}|^2/(\omega-\varepsilon_k^\alpha)$,
we start with the ansatz for the self-energy 
\begin{equation}
 \Sigma(\omega) = Un + \frac{ a\Sigma^{(2)}(\omega) }{ 
  1 - b\Sigma^{(2)}(\omega) } ,
     \label{eq:ansatz}
\end{equation}
where $n$ is the occupation number of the QD level
%
%\begin{displaymath}
% n = -\frac{i}{2\pi} \int_{-\infty}^\infty d\omega G^<(\omega)
%\end{displaymath}
%
which should be determined self-consistently, and
$\Sigma^{(2)}$ is the second order self-energy in $U$.
Note that each Green's function line in the $\Sigma^{(2)}(\omega)$
diagram is given by
\begin{equation}
 G_0(\omega) = \frac{ 1 }{ \omega-\varepsilon_0-Un
  - \sum_\alpha \Delta_\alpha(\omega) } \;.
\end{equation}

%and $\tilde{\varepsilon}_0$ is the renormalized QD level 
%which must be determined by imposing the Friedel sum rule:
%\begin{equation}
%$n = -\frac{1}{\pi} \delta(0)$.
%\end{equation}
%The phase shift created by the QD at the Fermi energy 
%is given by $\delta(0) = \mbox{\rm Im}[\ln G_\sigma(0)]$. 

The parameter $a$ in Eq.(\ref{eq:ansatz}) is determined from the 
condition that the self-energy has the exact behavior at high frequencies, 
and $b$ is determined from the atomic limit.~\cite{kajueter96} Both 
conditions lead to the expressions
\begin{equation}
 a = \frac{ n(1-n) }{ n_0(1-n_0) }
\end{equation} 
%
%($n_0=-i/(2\pi)\int_{-\infty}^\infty d\omega G_0^<(\omega)$),
%
and
\begin{equation}
 b = \frac{ (1-2n) }{ n_0(1-n_0) U } \;,
\end{equation}
where $n_0$ is a fictitious particle number obtained from
$G_0 (\omega)$. Hence, these two parameters give the correct weak and 
strong coupling limits in any region, including Kondo, charge fluctuation, 
and even-number site limit.

\begin{figure}[h]
\epsfxsize=3.5in
\epsffile{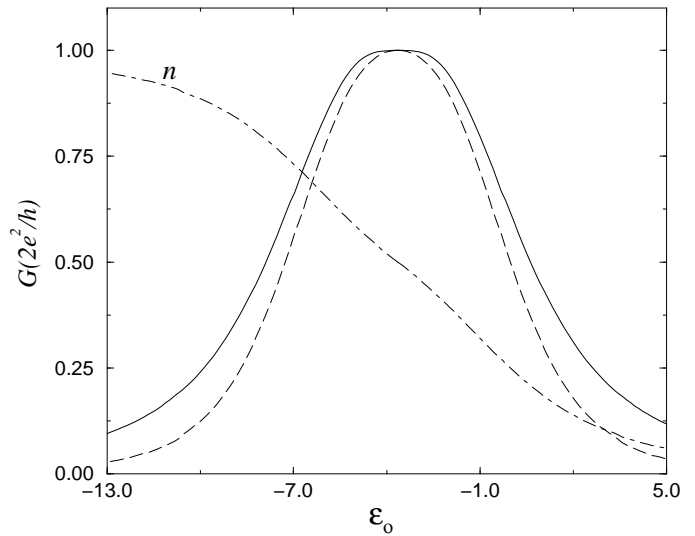}
  \caption{ Linear-response conductance at zero temperature as a function
   of the dot level position. The conductance is obtained from the DOS 
   (solid line) and from the Langreth's (exact) relation (dashed line). 
   The dot-dashed line denotes the occupation number of the QD. 
   $U=7.27\Gamma(0)$ and a parabolic form of $\Gamma(\omega)$ 
    is used in this and all the other figures.
	   }
  \label{fig1}
\end{figure}

The linear-response conductance $G = dI/dV|_{V=0}$ can be obtained from
the spectral density of states according to the current formula Eq.(\ref{eq:curr}).
At zero temperature, it reduces to
\begin{equation}
 G = \frac{2e^2}{h} \, 4\pi\tilde{\Gamma}(0)\rho_\sigma(0)  .
\end{equation}
In Fig.~\ref{fig1} we show the zero-temperature linear-response conductance 
as a function of the dot level for symmetric coupling 
$\Gamma_L(0)=\Gamma_R(0)$. 
For this and all the other figures we have chosen $U=7.27\Gamma(0)$. 
A parabolic form of $\Gamma(\omega)$ centred at $\omega=0$ with 
bandwidth $W=16.62\Gamma(0)$ is used and the energy scale in the 
$x$-axis is normalized to $\Gamma(0)$.
According to our results (see Fig.\ref{fig1}), $G\lesssim 2e^2/h$ 
in the Kondo limit, which implies the Kondo-assisted transmission.
At zero temperature, the linear-response conductance may also be deduced
from the Friedel-Langreth~\cite{langreth66} sum rule, which leads to the 
relation~\cite{ng88}
\begin{equation}
 G = \frac{2e^2}{h} \frac{4\Gamma_L(0)\Gamma_R(0)}
{(\Gamma_L(0)+\Gamma_R(0))^2}
   \sin^2{\pi n} \;. 
\label{eq:sumrule}
\end{equation}

In  Fig.~\ref{fig1} one can see that the conductance obtained from 
the density of states presents good agreement with those obtained from 
the sum rule. It is important to note that such agreement could not be 
achieved by other treatments such as the ordinary 2nd order perturbation 
theory or by the NCA.~\cite{nca}

\begin{figure}
\epsfxsize=3.5in
\epsffile{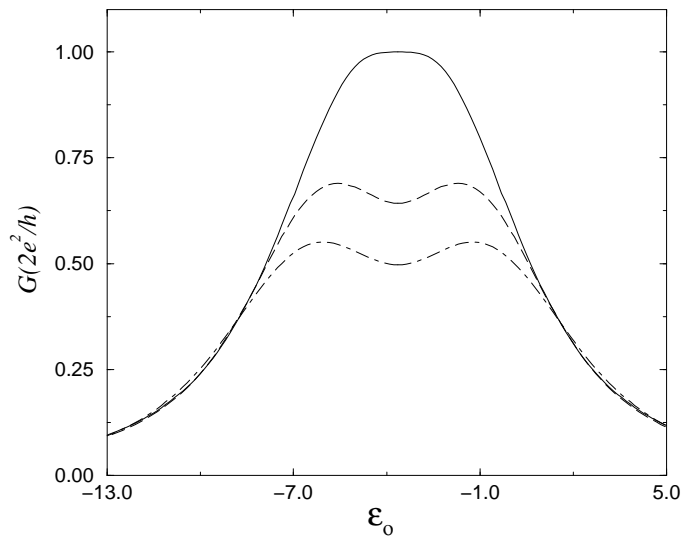}
  \caption{ Linear response conductance at $T=0$ (solid line), 
  $T=0.5\Gamma(0)$ (dashed line) and $T=\Gamma(0)$ 
  (dot-dashed line). The conductance is symmetric 
  around $\varepsilon_0=-U/2$ and it has minimum at this value
  for finite temperatures and a flat maximum at that point
  for $T=0$.
           }
  \label{fig2}
\end{figure}

The linear-response conductance at finite temperatures is displayed
in the Fig.~\ref{fig2}. The conductance shows a strong temperature
dependence in the Kondo limit, while it is almost temperature-independent
for $\varepsilon_0 > 0$ and $\varepsilon_0+U < 0$. 
The suppression of the conductance in the
Kondo limit at finite temperatures is related with the reduction
of the density of states near the Fermi level.
As a result, the conductance has two maxima at certain points in
$-U<\varepsilon_0<0$, and a minimum at $\varepsilon_0=-U/2$.
Note that the conductance is symmetric around $\varepsilon_0 = -U/2$
because of electron-hole symmetry of the model
and has a minimum value at this energy for $T\ne0$. 
To our knowledge, this is the first quantitative result which
shows the nonzero conductance minimum due to the Kondo effect.
Due to the Kondo resonance at finite temperature, the peak spacing
for an odd-number QD is smaller than that of an even-number QD.  
The peak spacing between two conductance maxima is reduced as the
temperature decreases, which can be attributed to the quantum fluctuation
of the number of electrons in the QD. These effects have been observed 
in recent experiments for the SET.~\cite{gordon,cronen98}
The asymmetry comes from a transition between 
Kondo and non-Kondo system and is reproduced in our
calculation. The Kondo effect shows up as a nonzero conductance
between the conductance peaks for the QD with odd number of electrons.
As the temperature increases, the minimum conductance at $\varepsilon_0
=-U/2$ decreases and  the positions of the maximum conductance 
approaches to the bare levels, which implies that the transport properties 
are similar to the ordinary resonant tunnelling at high temperature. 

\begin{figure}
\epsfxsize=3.5in
\epsffile{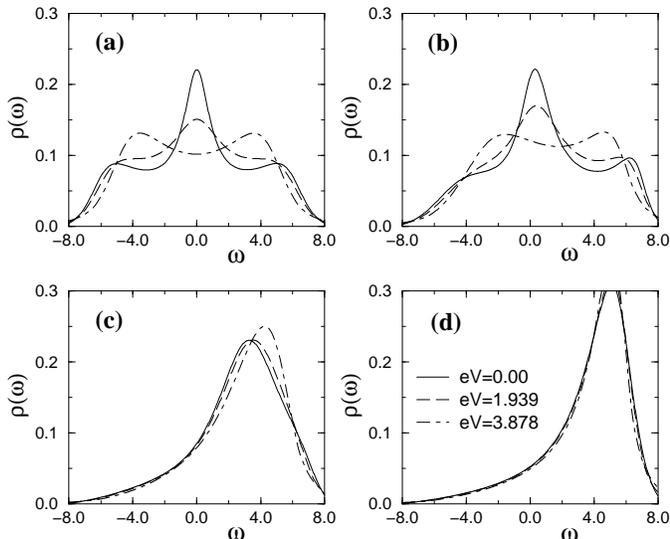}
  \caption{ Zero temperature single particle DOS of the QD level
   in the presence of the applied voltage. For (a) $\varepsilon_0 = -U/2$ 
   and (b) $\varepsilon_0=-0.286\,U$ the Kondo peaks are suppressed 
   by applying the voltage, while the shape of the spectral DOS remains 
   almost unchanged for non-Kondo systems, (c) $\varepsilon_0=0.19\,U$ 
   and (d) $\varepsilon_0=0.38\,U$.
           }
   \label{fig3}
\end{figure}
Let us now turn our attention to the non-equilibrium situation.
For simplicity, we assume a symmetric voltage drop, that is,
$\mu_L=-\mu_R=eV/2$.
Fig.~\ref{fig3} shows the spectral DOS of the QD at
non-equilibrium for different values of $\varepsilon_0$.
As one can easily find, the effects of the applied voltage are very
different for the Kondo systems ((a) and (b)) and for the non-Kondo 
systems~((c) and~(d)). In Fig.~\ref{fig3}~(a) and~(b) one can see that 
the resonance in the Fermi level is suppressed by applying the voltage,
because the external voltage produces inelastic scattering of 
quasi-particles located between the two different chemical potentials. 
For the symmetric case ($2\varepsilon_0+U=0$),
the spectral weight in the Fermi level is transfered to the satellite peaks. 
In the Fig.~\ref{fig3}~(b), we investigate the asymmetric case. Here, the 
reduction of  the Kondo peak is less pronounced than in  
the symmetric case.

%Note that the position of the single-particle peak moves toward lower energy 
%as the external voltage increases in this asymmetric case.

The DOS for $\varepsilon_0 > 0$, Fig.~\ref{fig3}~(c) and~(d), 
is insensitive to the external voltage. 
This is very different from the behavior of the Kondo system. 
In this limit, the transport properties are similar to the noninteracting 
system and the differential conductance has a weak dependence on 
the temperature and on the external voltage, as we discuss below.

\begin{figure}
\epsfxsize=3.5in
\epsffile{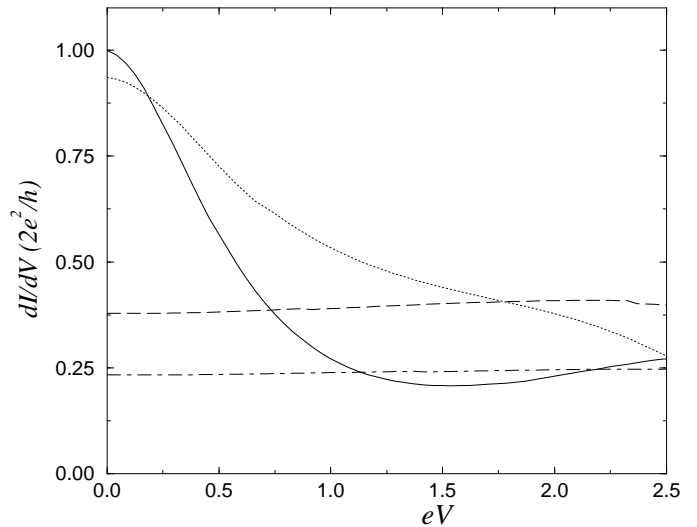}
  \caption{ Zero temperature differential conductance as a function of 
   the external voltage. Pronounced zero bias maxima are shown in the
   Kondo limit ($\varepsilon_0=-U/2$ (solid line), $-0.286\,U$ (dotted line)),
   while flat minima can
   be shown in the empty site limit ($\varepsilon_0=0.19\,U$ (dashed line),
   $0.38\,U$ (dot-dashed line)). 
           }
   \label{fig4}
\end{figure}

The differential conductance $dI/dV$ as a function of $V$
is displayed in Fig.~\ref{fig4} for several values of $\varepsilon_0$. 
The conductance is symmetric under bias reversal, since we considered 
a symmetric voltage drop as well as symmetric coupling for two tunnel barriers.
The conductance has a maximum at zero bias in the Kondo system,
because the Kondo resonance is reduced by the 
external voltage. This reduction is most pronounced in the symmetric 
case. This zero-bias maximum has been well known from previous 
calculations~\cite{hershfeld,meir,yeyati93} and observed in 
experiments.~\cite{gordon,cronen98,ralph94} 
The conductance for the symmetric case increases 
at larger voltage, which is related to the ordinary
resonance of the QD level with the leads.
In contrast, the conductance shows very weak zero-bias minimum
for the non-Kondo system ($\varepsilon_0 > 0$). This can be understood
from the behavior of the spectral DOS in the presence applied voltage
in Fig.~\ref{fig3}~(c) and~(d).  If the voltage dependence of the spectral
DOS is neglected, the conductance at zero temperature is given by 
\begin{equation}
 \frac{dI}{dV} = \frac{2e^2}{\hbar}
  \left[ \tilde{\Gamma}(eV/2)\rho_\sigma(eV/2)
       + \tilde{\Gamma}(-eV/2)\rho_\sigma(-eV/2) \right] .
 \label{eq:non-kondo}
\end{equation}
According to
the behaviour observed in Figs.~\ref{fig3}~(c) and~(d) as well as the 
Eq.~(\ref{eq:non-kondo}) one can conclude that the differential
conductance has a weak zero-bias minimum for a non-Kondo system.
The experimental results for the non-Kondo valley do not show 
a universal behavior.~\cite{cronen98} That is, both zero-bias minima
and maxima have been shown in the non-linear conductance.
This issue seems to be still open, while a pronounced zero-bias
minimum in the mixed valence limit
has been predicted in a previous study.~\cite{konig96}

In conclusion, we have studied resonant tunnelling through a quantum dot 
by means of a new perturbation treatment which correctly takes 
into account the weak and strong coupling limits. In our study, we
have found that the zero-temperature linear-response conductance agrees 
well with that obtained from the exact sum rule. 
At finite temperature the conductance has been shown
to have a nonzero minimum at the Kondo valley and the minimum 
conductance decreases with increasing temperature.
We have shown that the Kondo resonance is reduced by applying a
small voltage. This leads to a pronounced zero bias maximum 
of the differential conductance as a function of the voltage. 
On the other hand, the density of states is insensitive to
the voltage in a non-Kondo system, and accordingly weak zero-bias minima
occur in the differential conductance.

%The effects of the applied bias voltage
%in the single particle density of states and the differential conductances
%has been discussed both for the Kondo and the non-Kondo systems. 

%\acknowledgements
The authors wish to acknowledge P. Fulde for his hospitality during their stay
at the Max-Planck-Institut f\"ur  Physik Komplexer Systeme (MPI-PKS).
This work has been supported by the MPI-PKS. 
One of us (K.K) has also been supported in part by the KOSEF.

%%%%%%%%%%% References %%%%%%%%%%%%%

%
%\end{multicols}


\begin{references}
%
\bibitem[*]{kang} Electronic address : kckang@mpipks-dresden.mpg.de
%
\bibitem{kouwen97} L. P. Kouwenhoven {\em et al.},
 {\em Mesoscopic Electron Transport, Proceedings of the NATO Advanced
 Study Institute,} edited by L. L. Sohn, L. P. Kouwenhoven and G. 
 Sch\"on (Kluwer 1997).
%
\bibitem{glazman88} L. I. Glazman and M. E. Raikh, Pis'ma Zh. Eksp.
 Teor. Fiz. {\bf 47}, 378 (1988) [JETP Lett. {\bf 47}, 452 (1988)].
\bibitem{ng88} T. K. Ng and P. A. Lee, Phys. Rev. Lett. {\bf 61}, 1768 (1988).
\bibitem{hershfeld} S. Hershfeld, J. H. Davies and J. W. Wilkins,
 Phys. Rev. Lett. {\bf 67}, 3720 (1991); Phys. Rev. B {\bf 46}, 7046 (1992).
\bibitem{meir} Y. Meir, N. S. Wingreen and P. A. Lee, Phys. Rev.
 Lett. {\bf 70}, 2601 (1993);
 N. S. Wingreen and Y. Meir, Phys. Rev. B {\bf 49}, 11\,040 (1994).
\bibitem{yeyati93} A. L. Yeyati, A. Mart\'in-Rodero and F. Flores, Phys.
 Rev. Lett. {\bf 71}, 2991 (1993).
\bibitem{gordon} D. Goldhaber-Gordon, H. Shtrikman, D. Abush-Magder,
 U. Meirav and M. A. Kastner, Nature {\bf 391}, 156 (1998);
 D. Goldhaber-Gordon {\em et al.}, unpublished (cond-mat/9807233).
\bibitem{cronen98} S. M. Cronenwett, T. H. Oosterkamp, and L. P.
 Kouwenhoven, Science {\bf 281}, 540 (1998).
\bibitem{ralph94} D. C. Ralph and R. A. Buhrman, Phys. Rev. Lett.
 {\bf 72}, 3401 (1994).
\bibitem{kuramoto84} Y. Kuramoto and H. Kojima, Z. Phys. B {\bf 57},
 95 (1984); E. M\"uller-Hartmann, Z. Phys. B {\bf 57}, 281 (1984).
\bibitem{kajueter96} H. Kajueter and G. Kotliar, Phys. Rev. Lett. {\bf 77},
 131 (1996).
\bibitem{meir92} Y. Meir and N. S. Wingreen, Phys. Rev. Lett. {\bf 68},
 2512 (1992).
\bibitem{kang98} This formula can be equally applied to the 
 case of superconducting electrodes if the Andreev reflections are
 negligible due to large charging energy of the confined region. 
 See K. Kang, Phys. Rev. B {\bf 57}, 11\,891 (1998).
\bibitem{langreth66} D. C. Langreth, Phys. Rev. {\bf 150}, 516 (1966).
\bibitem{nca} The NCA is known to overestimate the Kondo-resonance.
 See the reference \onlinecite{meir}.
\bibitem{konig96} J. K\"onig, H. Schoeller and G. Sch\"on, Phys. Rev. Lett.
 {\bf 76}, 1715 (1996);
 J. K\"onig, J. Schmid, H. Schoeller and G. Sch\"on,
 Phys. Rev. B {\bf 54}, 16\,820 (1996).
%
\end{references}
\end{document}